\documentclass[aps,preprint,showpacs]{revtex4}
\usepackage{graphicx,bm}
\begin{document}
\newcommand{\half}{\frac{1}{2}}
\newcommand{\ith}{^{(i)}}
\newcommand{\im}{^{(i-1)}}
\newcommand{\gae}
{\,\hbox{\lower0.5ex\hbox{$\sim$}\llap{\raise0.5ex\hbox{$>$}}}\,}
\newcommand{\lae}
{\,\hbox{\lower0.5ex\hbox{$\sim$}\llap{\raise0.5ex\hbox{$<$}}}\,}
\newcommand{\Integer} {  \mbox{{\sf Z}} \hspace{-.43em} \mbox{{\sf Z}} }
\newcommand{\be}{\begin{equation}}
\newcommand{\ee}{\end{equation}}

\title{Tricritical O($n$) models in two dimensions}
\author{
Bernard Nienhuis~$^{1}$,
Wenan Guo~$^{2}$, and
Henk W.J. Bl\"ote~$^{3,4}$}
\affiliation{$^{1}$Instituut voor Theoretische Fysica,
Universiteit van Amsterdam, Valckenierstraat 65, The Netherlands}
\affiliation{$^{2}$Physics Department, Beijing Normal University,
Beijing 100875, P. R. China }
\affiliation{$^{3}$Faculty of Applied Sciences, Delft University of
Technology, P. O. Box 5046, 2600 GA Delft, The Netherlands}
\affiliation{$^{4}$Instituut Lorentz, Leiden University,
  P.O. Box 9506, 2300 RA Leiden, The Netherlands}

\date{\today} 
\begin{abstract}
  We show that the exactly solved low-temperature branch of the
  two-dimensional O($n$) model is equivalent with an O($n$) model with
  vacancies and a different value of $n$.  We present analytic results
  for several universal parameters of the latter model, which is
  identified as a tricritical point.  These results apply to the range
  $n \leq 3/2$, and include the exact tricritical point, the conformal
  anomaly and a number of scaling dimensions, among which the thermal
  and magnetic exponent, the exponent associated with crossover to
  ordinary critical behavior, and to tricritical behavior with cubic
  symmetry.  
  We describe the translation of the tricritical model in a Coulomb
  gas. The results are verified numerically by means of transfer-matrix
  calculations.  We use a generalized ADE model as an intermediary, and
  present the expression of the one-point distribution function in that
  language. The analytic calculations are done both for the square and
  the hexagonal lattice.

\end{abstract}
\maketitle 

\section {Introduction}
\label{intro}
The O($n$) model is defined in terms of $n$-component
spins on a lattice that interact in an isotropic way.
Graph expansion \cite{Stanley} of the partition integral of this
model leads to a weighted sum of graphs, in which every node is 
connected to its neighbors by an even numbers of bonds. 
In the resulting partition sum, the spin dimensionality $n$ is only a
parameter and may be varied continuously.
For a special choice of the model on the honeycomb lattice,
one thus derives a model of non-intersecting loops \cite{DMNS}.
Exact results for the universal parameters \cite{N} as a function
of $n$ were obtained for this O($n$) model for two cases, one
of them describing the critical point, and the other 
a special (see \cite{jrs}) case of the low-temperature phase. 
These results are described in the context of the Coulomb gas in a
review by Nienhuis \cite{BN2}.

These two cases of universal behavior were also found along two branches
of a square-lattice O($n$) model \cite{BNW,BN}, together with two different
branches describing the critical behavior that occurs when O($n$)
and Ising degrees of freedom on the square lattice display a joint
critical point. But it took a long time before an exact
description was also proposed \cite{GNB} for the {\em tricritical}
O($n$) universal parameters in two dimensions as a function of $n$. The
plausibility of this description follows from earlier identifications
of the fully packed O($n$) loop model with the critical Potts model, and
of the critical O($n$) model with the tricritical Potts model \cite{N}.
Therefore it seemed plausible that the tricritical O($n$) model
is associated with an even higher critical Potts model. Such a
model is known in the form of the tri-tricritical Potts model, 
for which some universal parameters are known \cite{NWB,KBN}.
Moreover, these universal parameters  were found to agree accurately
with numerical estimates of the conformal anomaly and some critical 
exponents of the tricritical O($n$) model on the honeycomb lattice.

In this paper we provide a more exact basis for this characterization
of the tricritical O($n$) universal parameters, by means of 
an exact mapping between an O($n$)-symmetric
spin model with vacancies, and an undiluted O($n'$) model which was
exactly solved in Refs.~\onlinecite{BNW} and \onlinecite{WPSN}.
This part is reported
in Sec.~\ref{smap}. Relations with other models, and the exact
solution are summarized in Sec.~\ref{exs}.  We include the mapping
on the Coulomb gas and derive exponents, including the one that 
is responsible for the  crossover to models with cubic symmetry. 
This part is presented
in Sec.~\ref{up}. In Sec.~\ref{num} we define the transfer matrix 
and apply it to confirm numerically some of the predicted exponents.
Results and consequences are discussed in Sec.~\ref{dis}.

\section{Mapping of the spin model on a solvable loop model}
\label{smap}
As a candidate system to display O($n$) tricriticality,
we choose a generalized version of the O($n)$ spin model on the
square lattice studied by Nienhuis and coworkers \cite{BN,BNW,WPSN}.
The $n$-component spins occupy the midpoints of the lattice edges.
They are denoted $\vec{s}_i$ where the index $i$ labels the
corresponding edge. The integration measure is normalized by
$\int d{\vec{s}}=1$ and the  mean length of the spins by 
$\int d{\vec{s}\,(\vec{s}\cdot \vec{s})}=n$.
The distribution is chosen isotropic,
i.e., in accordance with the O($n$) symmetry.
The model also includes face variables $t_j=0$ or 1, which sit on the
faces of the square lattice. The index $j$ labels the corresponding
face. The value $t_j=0$ corresponds with a `vacancy' which has the
effect of eliminating the interactions involving any of the four
adjacent spins. The vacancies thus introduce dilution and may thus be
expected to lead to a tricritical transition, analogous to that in 
the Potts model.

\begin{figure}
\begin{picture}(200,180)
\put(40,15){\includegraphics[scale=0.4]{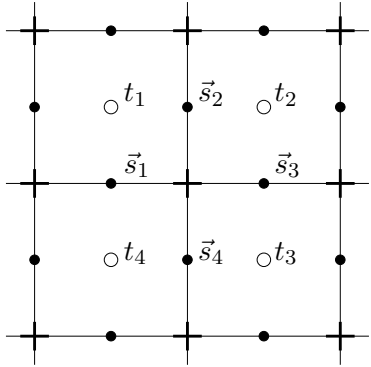}}
\put(85,115){$t_1$}
\put(113,115){$\vec{s}_2$}
\put(142,115){$t_2$}
\put(85,88){$\vec{s}_1$}
\put(142,88){$\vec{s}_3$}
\put(85,55){$t_4$}
\put(113,55){$\vec{s}_4$}
\put(142,55){$t_3$}
\end{picture}
\caption{Part of the square lattice with spin variables ($\bullet$),
and face variables ($\circ$) representing the vacancies. The vertices
are shown as {\bf +}. The figure includes the labeling of the variables
used in the definition of the local weight $W(k)$, when
applied to the central vertex in this figure.
}
\label{v1}
\end{figure}

\begin{figure}
\begin{picture}(450,220)
\put(0,15){\includegraphics[scale=0.75]{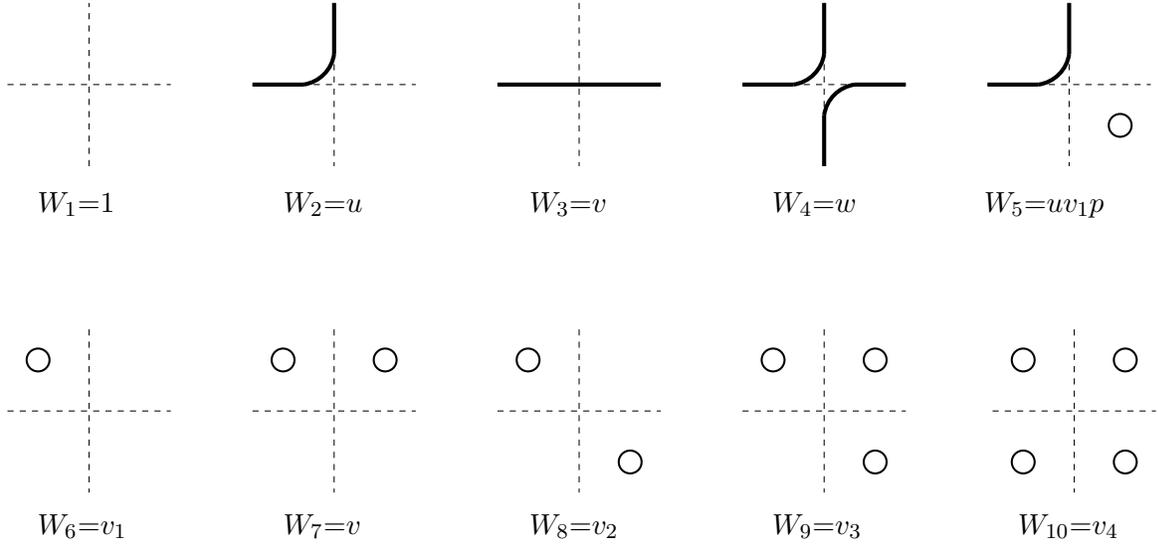}}
\put(12,122){$W_1$=$1$}
\put(105,122){$W_2$=$u$}
\put(198,122){$W_3$=$v$}
\put(290,122){$W_4$=$w$}
\put(370,122){$W_5$=$u v_1 p$}
\put(12,0){$W_6$=$v_1$}
\put(105,0){$W_7$=$v$}
\put(198,0){$W_8$=$v_2$}
\put(290,0){$W_9$=$v_3$}
\put(383,0){$W_{10}$=$v_4$}
\end{picture}
\caption{Vertex weights of the O($n$) model with vacancies. The solid
lines represent loop segments, and the open circles the vacancies on
the faces of the lattice. The presence of a vacancy implies that the
four surrounding edges are not visited by a loop. The absence of a
circle corresponds with an occupied face, whose edges may or may not
be visited by a loop. The same weights apply to rotated versions of
the vertices shown here. The spin variables, which sit on the middle
of the edges, are absent in the loop representation.
}
\label{v2}
\end{figure}

We write the partition sum as
\begin{equation}
Z_{\rm spin} =\left[ \prod_{{\rm faces}\;j} \sum_{t_j=0}^1 \right]
\left[ \prod_{{\rm edges}\;i} \int d{\vec{s}}_i \right] \prod_{{\rm vert}\;k} 
W(k)\,.
\label{Zspin}
\end{equation}
While the spins live on the edges of the lattice, and the vacancies on the
faces, the Boltzmann weight factorizes into 
factors $W(k)$ giving the interaction between all variables
incident on a vertex.
A part of the lattice is shown in
Fig.~\ref{v1}.
The local weight $W(k)$ is defined by
\begin{displaymath}
W(k)
\equiv \{ 1+u\sum_{i=1}^4 [(p+(1-p)t_{i+2}) t_it_{i+1}t_{i+3}  \;
\vec{s}_i \cdot \vec{s}_{i+1}]
\end{displaymath}
\begin{equation}
+t_1t_2t_3t_4 \sum_{i=1}^2[v \; \vec{s}_i \cdot \vec{s}_{i+2}
+ w (\vec{s}_i \cdot \vec{s}_{i+1})(\vec{s}_{i+2} \cdot \vec{s}_{i+3})]
\}V(t_1,t_2,t_3,t_4)
\label{wloc}
\end{equation}
where the indices of the $\vec{s}$ and $t$ variables are defined 
modulo 4. They describe the position of the variables in the local
configuration as specified in Fig.~\ref{v1} around the vertex $k$.
The vacancy weight $V(t_1,t_2,t_3,t_4)$ per vertex 
is specified by
\begin{displaymath}
V(t_1,t_2,t_3,t_4)\equiv \delta_{t_1+t_2+t_3+t_4,4} +
v_1\delta_{t_1+t_2+t_3+t_4,3}+ v_2 \delta_{t_1+t_2+t_3+t_4,2}
\delta_{t_1,t_3}
\end{displaymath}
\begin{equation}
+v \delta_{t_1+t_2+t_3+t_4,2} (1-\delta_{t_1,t_3})
+v_3 \delta_{t_1+t_2+t_3+t_4,1}+ v_4 \delta_{t_1+t_2+t_3+t_4,0}
\end{equation}
so that the index of $v_i$ indicates the number of vacancies.
The expansion in loop diagrams 
proceeds analogous to Ref.~\onlinecite{BN}, but
as a consequence of the added $t$ variables, the loops are
restricted to the edges that are not adjacent to a vacancy.
Furthermore, the term $p(1-t_{i+2})$ leads to an additional
potential for a loop segment that, with respect to a vertex, is
diagonally opposite to a vacancy. The loop expansion transforms the
partition function into 
\begin{equation}
Z_{\rm spin} = Z_{\rm loop}=
\left[ \prod_{j} \sum_{t_j=0}^1 \right]
 \sum_{\{{\mathcal L}\}|\{ t \}}
 n^{N_{\rm L}} \prod_{i=1}^{10} W_i^{N_i} 
\label{Zloop}
\end{equation}
where the second sum is on all configurations ${\mathcal L}$ of
closed loops, covering zero or more edges of the square lattice,
while avoiding edges adjacent to a vacancy. 
Every vertex is of one of ten types shown in Fig.~\ref{v2}.
The total number of
vertices of type $i$ is denoted  $N_i$, and the total number of
loops as $N_{\rm L}$.  The vertex weights $W_i$ are given in 
Fig.~\ref{v2}, in terms of the parameters 
that already appear in the spin representation of Eq.~(\ref{wloc}).

\section{Exact analysis }
\label{exs}
\subsection{Equivalence with the dense loop model}
\label{eqd}
Consider the exactly solved low-temperature branch of the O($n$) loop
model on the square lattice, named branch 2 in \cite{BN} and \cite{BNW}. 
Its partition sum, although of the form of Eq.~(\ref{Zloop}), is denoted 
$Z_{\rm dense~loop}$ referring to the relatively dense filling of the lattice
with loops. The weights can be parametrized in
terms of the angle $\theta \in [0,\pi/2]$.
Only the vertices labeled 1 to 4 in Fig.~\ref{v2}
have non-zero weight; their weights are specified as
\begin{eqnarray}
W_2  &=&  u= h(\theta) \sin( \theta  )  \nonumber \\
W_3  &=&  v= h(\theta) \sin(3\theta/4)  \label{clt}\\
W_4  &=&  w= h(\theta) \sin( \theta/4)  \nonumber 
\end{eqnarray}
with $h(\theta) \equiv 1/
[2\sin( \theta) \cos(3\theta/4) + \sin(3\theta/4)]$.
The weight of the loops (or the dimensionality of the spins) is
\begin{equation}
n'= -2 \cos(2 \theta)\, .
\end{equation}
We use $n'$ because we wish to reserve $n$ for another choice, in which
all weights $W_k$ are non-zero. 

\subsection{ADE models}\label{ADEfull}
\begin{figure}
\begin{picture}(250,180)
\put(15,15){\includegraphics[scale=0.5]{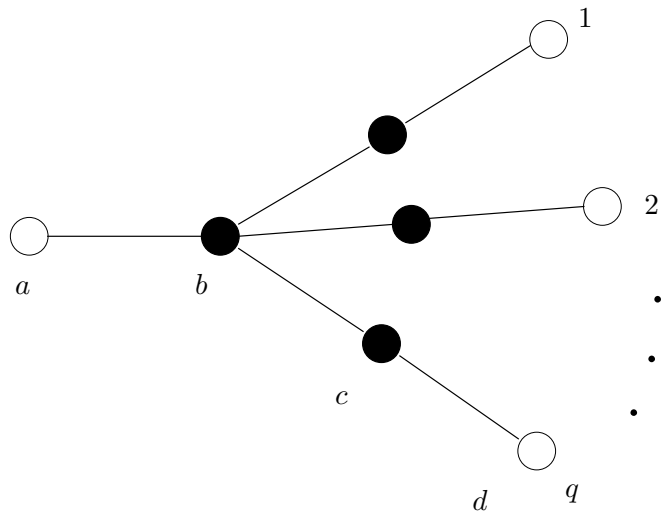}}
\put(17,82){$a$}
\put(85,82){$b$}
\put(138,40){$c$}
\put(190,0){$d$}
\put(230,183){$1$}
\put(255,112){$2$}
\put(225,5){$q$}
\end{picture}
\caption{The adjacency diagram ${\mathcal A}$ having $q$ branches. 
For $q=0, 1$ and 2, it reduces to the diagrams A$_2$, A$_4$ and  E$_6$
respectively.  
\label{ade} }
\end{figure}

Here we construct an alternative representation of the loop model
partition sum. Following Pasquier~\cite{Pasq1,Pasq2},
the loops are interpreted  as domain
walls in a configuration of discrete variables living on the faces of
the lattice.  
These variables take values,
corresponding to the nodes of a graph ${\mathcal A}$ called  the
adjacency diagram.
In this paper we consider the family of graphs  shown in
Fig.~\ref{ade}, but the discussion in this section is general and the
figure can be seen as an example.
We call this model an ADE model after the classification of adjacency
diagrams. 
Neighboring faces not separated by a loop carry the same value.  If they
are separated by a loop, their values are {\em adjacent} in ${\mathcal
  A}$ (hence its name). 
                                                                               
Associated to ${\mathcal A}$ we introduce the adjacency matrix ${\bf A}$
with elements $A_{ij}$, where $i$ and $j$ represent nodes of ${\mathcal
A}$.  The elements are defined as $A_{ij}=1$ if $i$ and $j$ are
adjacent, and $A_{ij}=0$ otherwise. Of the $2(1+q)$ eigenvectors of
${\bf A}$, four are symmetric under permutation of the $q$ branches.  In
the symmetric subspace the eigenvector equation can be written as
that for the right hand eigenvector of the $4 \times 4$ matrix ${\tilde
A}_{ij}$
\begin{equation}
\tilde{\bf A }=
\left (
\begin{array} {cccc}
0 & ~1 & ~0 & ~0\\
1 & ~0 & ~q & ~0\\
0 & ~1 & ~0 & ~1\\
0 & ~0 & ~1 & ~0
\end{array}
\right ) \, .
\label{adjm}
\end{equation}
The first two elements of the eigenvectors of $\tilde{A}$ correspond to
  the nodes $a$ and $b$ respectively, and the last two with the nodes
  of type $c$ and $d$ respectively (see Fig.~\ref{ade}). 
We choose the number of branches of the diagram ${\cal A}$ to be 
$q=(n'-1/n')^2$, so that 
the symmetric eigenvalues $\Delta_{\mu}$ of $\tilde{A}$ are given by
\begin{eqnarray}
\Delta_0 &=&   n' \nonumber \\
\Delta_1 &=& 1/n' \nonumber \\
\Delta_2 &=&-1/n' \label{ddef}\\
\Delta_3 &=& - n' \, , \nonumber 
\end{eqnarray}
and for each of these eigenvalues, the elements of the 
corresponding eigenvectors are
\begin{equation}
(S^{\mu}_a,S^{\mu}_b,S^{\mu}_c,S^{\mu}_d)=
(\Delta_{\mu}-1/\Delta_{\mu}\,,\;
\Delta_{\mu}(\Delta_{\mu}-1/\Delta_{\mu})\,,\; 
\Delta_{\mu}\, ,\; 1) \, .
\label{eivec}
\end{equation}
Besides these there are $2(q-1)$ eigenvectors antisymmetric for
interchange of two of the branches of ${\mathcal A}$.
They have eigenvalues 
\begin{eqnarray}
\Delta_4 &=& 1   \label{ddefc}\\
\Delta_5 &=&-1   \, , \nonumber
\end{eqnarray}
each $(q-1)$-fold degenerate.

Of the eigenvectors of ${\bf A}$, generally denoted as $S^{\mu}_j$, we omit
the upper index for the case $\mu=0$. For $n'>1$ this is the
Perron-Frobenius eigenvector, with eigenvalue $\Delta_0=n'>\Delta_{\mu}$
for $\mu=1,2,3$.

We write the weight of the corresponding ADE model in terms of a product
of local weight factors as
\begin{equation}
W_{\rm ADE}=\prod_{{\rm vert}\;k} W(k)
\prod_{\rm turns}
A_{ij}\left(\frac{S_i}{S_j}\right)^{\gamma_{\rm bend}/2\pi} \, .
\label{WADE}
\end{equation}
There is a factor $W(k)$ for each vertex depending only on the local
configuration of domain walls, and a factor for each turn of the domain
wall which also depends on the states of the faces on the inside ($i$)
and outside ($j$) of the loop.   The bending
angles $\gamma_{\rm bend}$ are counted positive where the loop bends inwards, 
so that the sum of the bending angles along a loop is $+2\pi$. 
Thus the weight of an entire closed domain wall is
\begin{equation}
W_{\rm loop} = A_{i,j} {S_i\over S_j}\,,
\label{ADEloop}
\end{equation}
and still depends on the state $i$ inside and $j$ outside the domain wall. 
When for a fixed configuration of domain walls the
sum over compatible state configurations is performed, each closed domain
wall thus contributes a factor equal to the largest eigenvalue of ${\bf A}$,
i.e. $n'$, just as in the O($n'$) loop model. 
This confirms that the partition function of the ADE model
satisfies
\begin{equation}
Z_{\rm ADE} \equiv \sum_{\cal G} \sum_{\{s\}|{\cal G}} W_{\rm ADE}
=Z_{\rm dense~loop} \, ,
\label{ZADE}
\end{equation}
where the first sum is on all loop configurations and the second one on
the configurations of ADE variables compatible with the loops.  
The suffix of the right hand
side emphasizes that the loop model is in the dense phase, and does not
permit vacancies, which will be introduced later. 
Since in $Z_{\rm dense~loop}$ the
variable $n'$ and therefore $q$ only enter as parameters we may vary them
continuously, thus representing the continuous $n'$-weight loop model. 

When the models live on a torus rather than in the plane, there may be
loops that wind the torus.  In the loop model these typically have 
the same weight
$n'$ as the contractible loops.  However, in the ADE model, the
corresponding domain walls have a net bending angle equal to zero.  
This implies
that they carry the weight $A_{j,k}$ rather than $A_{j,k} S_j/S_k$.
The summation over the states of the domains then reduces to taking
the trace of a power of ${\bf A}$, equal to the number of
non-contractible loops.
The result is that all the winding loops have the same weight, equal to
an eigenvalue of ${\bf A}$, which should then be summed over these eigenvalues.
We conclude that the ADE model on a torus corresponds to a loop model in
which the winding loops receive special treatment. Or alternatively
the loop model partition sum with all loops weighted equally is the
largest sector of the ADE transfer matrix.

\subsection{Correlation functions}\label{corfun}
We will now calculate the one-point distribution (1PD) $P(k)$ of the
ADE model, i.e., the probability that a face is in state $k$.
Consider a loop well inside a large lattice.  We assume that the 1PD is
unaffected by the presence of the loop (or any other loop).   In other words
we assume that the 1PD conditional on the presence of a loop is the same
as the unconditional 1PD.   That this is plausible follows from the
calculation of the partition sum above: the contribution to the
partition sum of a particular domain is independent 
of the domains it is contained in, and it is independent of all the
domains it contains, once the state of these domains has been summed over.

The conditional probability $P(k|j)$ that the inside domain of a loop is
in state $k$, provided the outside domain is in a given state $j$, is
determined by Eq.~(\ref{ADEloop}) as $P(k|j)=A_{k,j}S_k/(n'\,S_j )$. 
Thus we find the joint probability $P(k,j)$ that the outside of a loop
is in state $j$ {\em and} its inside in state $k$ as
\begin{equation}
P(k,j) = P(j)P(k|j)=P(j) A_{k,j} \, \frac{S_k}{n'\, S_j}\,.
\end{equation}
Summation on $j$ now yields the probability that the inside domain
is in state $k$, which should be equal to $P(k)$:
\begin{equation}
\sum_j P(k,j) = P(k)\;.
\end{equation}
Using the symmetry of ${\bf A}$, one finds 
the unique (normalized) solution to this consistency condition as 
\begin{equation}
P(k) = \frac{S_k^2}{\sum_j \left(S_j^2\right)} \, .
\label{pk}
\end{equation}

An approach alternative to the condition that the loop considered is
well inside a large lattice, is to consider a bounded lattice of
arbitrary size, with the faces on the boundary all in the same state,
with Eq.~(\ref{pk}) as the probability distribution for that state, the
ideal fixed boundary condition.  Then by induction the same distribution
holds for the domains separated from the boundary by one domain wall,
and so on recursively 
to the innermost domains.  It is then assumed that in the
thermodynamic limit the boundary condition should not matter, well away
from the boundary.

Consider the function $S^{\mu}_k/S_k$, i.e., the ratio of an
arbitrary eigenvector $S^{\mu}$ and the Perron-Frobenius eigenvector $S$.  
If this function is part of a correlation function 
$\langle\cdots  S^{\mu}_k/S_k \cdots\rangle$, where $k$ is the state of
a given face, it  effectively 
changes the weight of the loops {\em surrounding the face}.  
This is easily seen in the expression (\ref{ADEloop}): the
factor  $S^{\mu}_k/S_k$ replaces the numerator by $S^{\mu}_k$,  so that
the weight of the loop becomes that of the corresponding eigenvalue,
$\Delta_{\mu}$, as long as they do not surround other operator insertions. 
We will call these functions weight-changing operators.

A more interesting result~\cite{Pasq3} comes from the two-point function 
\be \left\langle {S^{\mu}_j\over S_j}\; {S^{\nu}_k\over S_k} \right\rangle\;,\ee
$j$ and $k$ being
the state of two arbitrary faces. The weights of the 
loops surrounding either of these
faces but not the other is changed into the respective eigenvalues
$\Delta_{\mu}$ and $\Delta_{\nu}$.
corresponding to the eigenvectors $S^{\mu}$ and $S^{\nu}$. 
Now consider the innermost domain wall that surrounds both faces. 
After the states of the domains nested inside it are summed over, the weight
governing the state of the final domain 
is \be {S^{\mu}_j S^{\nu}_j\over S_j^2} A_{j,k}\;, \ee
where $k$ is the state of the surrounding domain.
This can be expanded as a linear combination of all eigenvectors:
\be  S^{\mu}_j S^{\nu}_j =  S_j \sum_{\kappa}C_{\mu\;\nu}^{\kappa}
S^{\kappa}_j\,, \ee
where, provided the eigenvectors are normalized,
\be C_{\mu\;\nu}^{\kappa} = \sum_j {S^{\mu}_j S^{\nu}_j S^{\kappa}_j\over S_j}. \ee
Apparently the combination of two operators labeled $\mu$ and $\nu$
look from a distance like a linear combination of operators $\kappa$.

These structure constants of the operator product expansion, or fusion
rules, may readily be
calculated explicitly for the diagrams in Fig.~\ref{ade}, but here we
only note that they are symmetric in $\mu$, $\nu$ and $\kappa$ and that
they vanish if one of the indices corresponds with the largest
eigenvalue, and the others two differ. This implies that the
two-point correlation function of two different weight-changing 
operators vanishes in the thermodynamic limit.
Obviously these fusion rules may be used just as well in
correlation functions of more than two operators.

\subsection{Equivalence with the O($n$) model with vacancies}
\label{onvac}
We interpret the extrema of ${\mathcal A}$ as vacancies, 
that is  the nodes $a$ and those of type $d$.
Thus, there may be $q+1$ types of vacancies in an ADE configuration. 
However, in any configuration the type of each vacancy is fully
determined by the neighboring domains. Thus, it is sufficient to
specify the $q+1$ states for the non-vacant domains in order to
fully describe an the ADE configuration (with the exception of
the completely vacant state).

Now we identify the domain walls between domains in state $b$ and in
states of type $c$ as loops, so that we have a loop model with vacant 
faces.  
Note that for any given configuration of vacancies this loop model is
much like that described in Sec.~\ref{ADEfull}, but on a restricted
lattice, from which the vacant faces are omitted, and with a reduced
adjacency diagram, of only the full nodes in Fig.~\ref{ade}.
Note that the eigenvectors of the reduced adjacency matrix, up to
normalization, are the same as the eigenvectors of the total adjacency
matrix, restricted to the nodes $b$ and those of type $c$.
The eigenvalues are $\pm\sqrt{q}$. 
Following the arguments used before, we obtain a
loop model with loop weight $n = \sqrt{q}$, with weights that follow
from the original ADE model with the complete adjacency diagram and the
entire lattice.
Thus the successive transformations are
\begin{equation}
Z_{\rm dense~loop} \to Z_{\rm ADE} \to Z_{\rm loop+vac.}\;,
\label{map}
\end{equation}
in which the right hand side is the partition sum of a model with loops
and vacancies.
The weights of this loop model are, with reference to Fig.~\ref{v2},
given by
\begin{eqnarray}
W_1~  &=& 1   \nonumber \\
W_2~  &=& u \nonumber \\
W_3~  &=& v \nonumber \\
W_4~  &=& w \nonumber \\
W_5~  &=& w (S_a/S_b)^{1/4}=w(S_d/S_c)^{1/4}=w(n')^{-1/4}    \nonumber\\
W_6~  &=& u (S_a/S_b)^{1/4}=u(S_d/S_c)^{1/4}=u(n')^{-1/4}    \nonumber\\
W_7~  &=& v \nonumber \\
W_8~  &=& w [(S_a/S_b)^{1/2}+(S_b/S_a)^{1/2}]=
w [(S_d/S_c)^{1/2}+(S_c/S_d)^{1/2}]=w [(n')^{-1/2}+(n')^{1/2}] \nonumber\\
W_9~  &=& u (S_b/S_a)^{1/4}= u (S_c/S_d)^{1/4}=u (n')^{1/4}  \nonumber\\
W_{10} &=& 1
\label{wvwv}
\end{eqnarray}
respectively. The two terms in $W_8$ arise from the two orientations
of the type-4 vertex of the O($n'$) loop model. 

We note that the weights are completely given by the configuration of
loops and vacancies, irrespective of the type of vacancy.
Furthermore, any configuration of loops and vacancies consisting of the
local vertices in Fig.~\ref{v2} is  possible for the adjacency diagram
in Fig.~\ref{ade}.  These properties are not generic 
for any adjacency diagram, and are the basis of our choice of the
diagram in Fig.~\ref{ade}, together with the fact that is contains a
continuously variable parameter controlling the eigenvalues of its
adjacency matrix.

\subsection{The O($n$) model on the honeycomb lattice}
A similar O($n$) spin model with vacancies on the faces can be
defined on the honeycomb lattice, see e.g. Ref.~\onlinecite{GNB}.
Here we also include interactions between the vacancies, described 
by the three vertex weights $v_1$, $v_2$, and $v_3$ where the indices
show the number of vacancies adjacent to the vertex.

\begin{figure}
\begin{picture}(450,110)
\put(10,25){\includegraphics[scale=0.45]{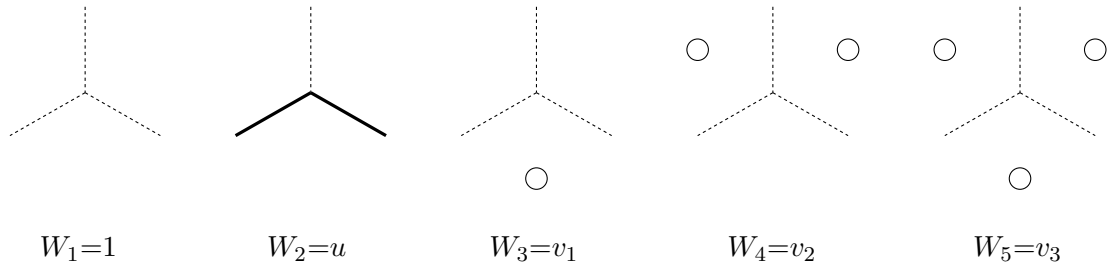}}
\put(22,0){$W_1$=$1$}
\put(108,0){$W_2$=$u$}
\put(192,0){$W_3$=$v_1$}
\put(282,0){$W_4$=$v_2$}
\put(375,0){$W_5$=$v_3$}
\end{picture}
\caption{Vertex weights of the honeycomb O($n$) model with vacancies
on the faces. Solid lines represent loop segments, and circles the
vacancies on the elementary hexagons. A vacancy excludes the
six surrounding edges to be visited by a loop. Rotated versions of
these vertices have the same weights. The spin variables, which sit on
the vertices, play no part in the loop representation.
}
\label{hcvw}
\end{figure}

The transformation into a loop model partition sum proceeds the same
as for the square lattice, and leads to the form of Eq.~(\ref{Zloop})
but with only five independent vertices. They 
are shown in Fig.~\ref{hcvw}, together with their weights.
For the simplified case $v_1=v_2=v_3=0$ without vacancies, and with the
special choice \cite{BN82}
\begin{equation}
u = 1/\sqrt{2 \pm \sqrt{2-n'}} \, ,
\label{crpt}
\end{equation}
(which is different from that for the square lattice),
this model is solvable \cite{BN82,Baxh,BB}. The high loop density branch
of this model corresponds with the minus sign. 

The mapping of the low-temperature O($n'$) model on the dilute O($n$)
model can be performed analogously for the honeycomb lattice,
and leads to the following vertex weights for the model with vacancies:
\begin{eqnarray}
W_1  &=& 1   \nonumber \\
W_2  &=& u \nonumber \\
W_3  &=& u (S_a/S_b)^{1/6}=u(S_d/S_c)^{1/6}=u(n')^{-1/6}  \nonumber\\
W_4  &=& u (S_b/S_a)^{1/6}=u(S_c/S_d)^{1/6}=u(n')^{ 1/6}  \nonumber\\
W_5  &=& 1   \, . \nonumber
\label{wvh}
\end{eqnarray}
The adjacency matrix, and thus its eigenvector components, are the same
as for the square lattice model, Eq.~(\ref{eivec}). Also the loop weight
\begin{equation}
n = n'-\frac{1}{n'}\,,
\label{nn'}
\end{equation}
is the same.

\subsection{Interpretation }
It remains to be shown that the constructed model of loops and vacancies
is a tricritical O($n$) model. 
To this purpose we obtain its conformal anomaly $c$ via the equivalence
with the low-temperature O($n$) model of Eq.~(\ref{clt}),
for which \cite{BN,BNW}
\begin{equation}
c=1-\frac{6(1-g)^2}{g}\,, \mbox{\hspace{4mm}}
2 \cos (\pi g)=-n' \,, \mbox{\hspace{4mm}}
0\leq g \leq 1 \,, \mbox{\hspace{4mm}}
n=n'-\frac{1}{n'} \,.
\label{cn1}
\end{equation}
This combination of $c$ and $n$ does not agree with the known
critical and low-temperature O($n$) universality classes \cite{N}.
Since it is known that the introduction of vacancies can lead to
tricriticality, this already suggests that the model defined by
Eqs.~(\ref{Zloop}), (\ref{wvwv}) and (\ref{nn'}) is 
tricritical. Further justification will be given below.

\section{Universal properties and mapping on the Coulomb gas}
\label{up}
\subsection{The conformal anomaly}
Eq.~(\ref{cn1}) shows that the model
with vacancies derived above does not fit the critical and
low-temperature O($n$) universality classes \cite{N}. It does,
however, precisely match the result for $c$ inferred in
Ref.~\onlinecite{GNB} for the tricritical O($n$) model, namely
\begin{equation}
c=1-\frac{6}{m(m+1)}\,, \mbox{\hspace{4mm}}
2 \cos \frac{\pi}{m+1}=\Delta\,, \mbox{\hspace{4mm}}
m \geq 1 \,, \mbox{\hspace{4mm}}
n=\Delta -\frac{1}{\Delta} \,.
\label{cn2}
\end{equation}
The parametrizations of $n$ in Eqs.~(\ref{cn1}) and (\ref{cn2})
imply that $n'=\Delta$.
This provides a confirmation that the O($n$) loop model defined
by the vertex weights of Eq.~(\ref{wvwv}) and the loop weight of
Eq.~(\ref{nn'}) is a tricritical O($n$) model. 
For the Ising model $n=1$ and the self-avoiding walk $n=0$ this
proposal reproduces the the known
tricritical values of $c=7/10$ and $c=0$ respectively. 

\subsection{Relation with the Coulomb gas}
The Coulomb gas offers a powerful tool to calculate critical
exponents, provided the Coulomb gas coupling constant $g$ is
known for the model under investigation. The coupling constant is
related \cite{BCN,Aff} to the conformal anomaly $c$
according to Eq.~(\ref{cn1}).
In combination with Eq.~(\ref{cn2}) this yields $g$ as a function
of $m$ which is two-valued. For the tricritical O($n$) model one
has $g=m/(m+1)$, in agreement with the conformal classification
of the tricritical Ising ($n=1$) model.
In the Coulomb gas language, the conformal anomaly is represented
by means of a pair of electric charges  $e_0$ as follows:
\begin{equation}
c= 1-\frac{6e_0^2}{g} \, , \quad e_0=1-g \,,\quad n' = -2 \cos\pi g\,.
\end{equation}
Combination with another pair of charges $\pm e_{\mu}$ yields a
scaling dimension
\begin{equation}
X_j = \frac{e_{\mu}^2 - e_0^2}{2g}  \, ,
\label{thcg}
\end{equation}
which reproduces the dimensions listed in Refs.~\onlinecite{GNB} 
and \onlinecite{NWB} for the tricritical O($n$) universality class
for charges $e_{\mu}$ according to \cite{BN2} 
\begin{equation}
\cos (\pi e_{\mu})= \frac{\Delta_{\mu}}{2}\, .
\label{ej}
\end{equation}
The $\Delta_{\mu}$ follow from the eigenvalues of the adjacency matrix
\cite{Pasq1,Pasq2}, which is the same as that used for the 
tri-tricritical Potts model \cite{NWB}.
Six eigenvalues were already listed in Ref.~\onlinecite{NWB}, and in
Eqs.~(\ref{ddef}) and (\ref{ddefc}).
The leading scaling  dimensions  follow as
\begin{eqnarray}
X_1 &=& \frac{ (1-t)^2
         -(1-g)^2} {2g}                      \nonumber \\
X_2 &=& \frac{ t^2
         -(1-g)^2} {2g}                      \nonumber \\
X_3 &=& 1-\frac{1}{2g}                  \label{Xdef}   \\
X_4 &=& 1-\frac{4}{9g}  -\frac{g}{2}         \nonumber \\
X_5 &=& 1-\frac{5}{18g} -\frac{g}{2}    \, , \nonumber 
\end{eqnarray}
with
\begin{equation}
t = \frac{1}{\pi}\,\arccos\frac{1}{2\cos(\pi g)} \, .
\label{tdef}
\end{equation}
These exponents are associated with  the weight-changing
operators $S^{\mu}_k/S_k$, discussed in Sec.~\ref{corfun}.
This implies, see Eq.~(\ref{eivec}),
that the  $X_2$ is associated with the enhancement (or
suppression) of vacancies: its eigenvector has a different signs in the
vacant and non-vacant states. It follows that $X_2$ is one of the thermal
exponents, leading or otherwise.
The exponent $X_1$ belongs to the operator that breaks the
pseudo-symmetry between the left and right hand side of ${\mathcal A}$.
It changes the weight of the loops from
$n$ to $-n$, but is not meaningful in the local O($n$)
spin version of the loop model. One can arrange this effect, however, by
an operator that terminates a seam across which  the interactions
have the opposite sign. It will appear later as the interface exponent.
The exponent $X_3$ combines the effect of the prior two operators.
Finally the the operator associated with the exponent $X_4$
breaks the symmetry between the $q$ branches, while that of $X_5$
combines this with the enhancement of vacancies. Again, these operators
only show up in models where the nodes of ${\mathcal A}$ are meaningful.
In our O($n$) models, the diagram is only used to give the proper weights
to various configurations.

In addition to the dimensions describing singularities associated
directly with the ADE model,
we consider exponents describing crossover phenomena due to
a possible perturbation of the O($n$) symmetry. Such perturbations
can, in the Coulomb gas language, be associated with pair of electric
charges $1-g$ combined with a pair of magnetic charges whose magnitude
depends on the type of perturbation \cite{BN2}. A pair of magnetic
charges $\pm k$, associated with the defect of $2k$ lines coming
together, then corresponds with a scaling dimension
\begin{equation}
X_{p,k} = 1-\frac{1}{2g}+\frac{g(k^2-1)}{2}
\label{wm}
\end{equation}

\subsection{Specific exponents}
\noindent
{\em The magnetic exponent}\\
In the O($n$) loop model, the magnetic correlation function is
represented by configurations containing a single loop segment
connecting the correlated points. In the Coulomb gas language
it corresponds with magnetic charge $k=1/2$. Then,
Eq.~(\ref{wm}) yields
\begin{equation}
X_h=1-\frac{1}{2g}-\frac{3g}{8} \, .
\label{xh}
\end{equation}
This is in agreement with an earlier conjecture \cite{js} and with
numerical results for a related model with vacancies on the honeycomb
lattice \cite{GNB}. After the mapping onto the model with vacancies, the
defects, i.e., the end points of the loop segment, can only sit on
non-vacant sites,  and  not in the regions occupied by vacancies.
But otherwise, the defects have the same physical effect, and therefore
the same exponent (\ref{xh}).
Therefore the magnetic exponent of the tricritical O($n$) model
is the same as that of the low-temperature O($n'$) model.

\noindent
{\em The temperature exponent}\\
Above we already identified the exponent $X_2$ as a thermal exponent.
This agrees with identification  on the basis of numerical evidence,
in \cite{GNB} as, in fact, the leading thermal exponent.
Effects described by this exponent are absent in the thermal
properties of the low-temperature O($n'$) model. This fits well in the
interpretation \cite{N} that the latter model is confined to the critical
subspace of a larger parameter space. Such `unphysical' O($n'$)
exponents are however known to reappear in correlations on dilute
O($n$) loop configurations \cite{Dengea}.
We further remark that the thermal exponent of the O($n'$) model,
associated with the suppression and enhancement of loops, should be
another thermal exponent; its value, $4/g-2$, however indicates that
it is irrelevant.
Another option to find a second thermal exponent, comes from the
observation that the value of $t$ in Eq.~(\ref{tdef}) is only the
smallest solution of the inverse cosine. The next leading exponent is
obtained by replacing $t$ by $2-t$ in $X_2$, so that
\begin{equation}
X_2 \to X_{t2} = \frac{ (2-t)^2 -(1-g)^2} {2g}
\end{equation}

\noindent
{\em Other exponents}\\
The introduction of a `seam', i.e., a row of antiferromagnetic bonds,
such that the bonds are perpendicular to the row, leads to a change
of the partition sum described by an `interface dimension $X_m$.
While in  Ref.~\onlinecite{GNB} the identification $X_m=X_1$ was made on
the basis of numerical evidence,
here we can make the identification by inspection of the corresponding
operator. A seam along the length of the cylinder
changes the sign of the non-contractible loops.
That corresponds precisely with
$\Delta_1=1/n'$ in  Eq.~(\ref{ddef}) because inversion of $n'$
results in a change of sign of $n$.

In the O($n$) spin model, other interface exponents can be constructed
by the introduction of a cut across which the spin $\vec{s}$ is identified
with ${\bf R}\cdot \vec{s}$, where ${\bf R}$ is an operator in the
orthogonal group
O($n$). The weight of the loops crossing this cut is then equal to
$\int d{\vec{s}\,(\vec{s}\cdot{\bf R}\cdot \vec{s})}$.
By varying ${\bf R}$ this can take any value.
The corresponding exponent is obtained by the relation 
$\Delta-1/\Delta = \int d\vec{s}\;\vec{s}\cdot {\bf R}\cdot \vec{s}$, and 
Eqs.~(\ref{thcg}) and (\ref{ej}).

A cubic perturbation of the O($n$) symmetry can be represented by
magnetic charges $k=\pm 2$ in Eq.~(\ref{wm}) \cite{N,DS} which
yields
\begin{equation}
X_{p,2} = 1-\frac{1}{2g}+\frac{3g}{2} \, .
\label{cc}
\end{equation}
It is the exponent that describes the crossover when a cubic symmetry
breaking is introduced.
Also when intersections between the loops are permitted, this exponent
governs the crossover to another universality class. This indicates the
fact that the results in this paper are applicable exclusively when
intersections are prevented, by the specific choice of the Hamiltonian.

\section{Numerical verification}
\label{num}
For the construction of the transfer matrix we choose the usual geometry
of a model wrapped on a cylinder, such that one of the lattice edge
directions runs parallel to the axis of the cylinder.
The transfer-matrix method used here is based on that of Ref.~\onlinecite{BN},
including the sparse-matrix composition. The main modification is the
generalization of the set of connectivities used in  Ref.~\onlinecite{BN} 
to include the specification of the vacancy variables on the faces. 

\subsection{Enumeration of the connectivities}
We consider the model of Eq.~(\ref{Zloop}) on a cylinder with a
circumference of $L$ lattice units. The cylinder is has an open end such
that there are $L$ external edges, which may or may not be covered by
segments of incomplete loops. The connectivity specifies the following
information: (1) which of the faces at the end of the cylinder carry
vacancies; (2) the way in which pairs of covered external edges are
connected by incomplete loops of ${\cal L}$. These connectivities are
subject to the restriction that the loop segments cannot be adjacent to
a vacancy. Each connectivity can be fully specified by a row of integers
$(i_1,i_2,\cdots,i_L)$ such that
\[    \left \{ \begin{array}{ll}
    i_l=i_m>0  & \mbox{if and only if edge $l$ is connected to edge $m$}\\
    i_k=0      & \mbox{if and only if edge $k$ is not visited by a loop
                       segment  and the}\\
               & \mbox{ face to the right of $k$ is occupied}\\
    i_n=-1     & \mbox{if and only if face to the right is vacant.} \\
     \end{array}
       \right.   \]
The positions of the vacancies can simply be coded by means of an $L$-bit
binary number $(p_1,p_2,\cdots,p_L)$ with value 
${\beta}=\Sigma_{k=1}^{L} p_k2^{k-1}+1$. For a given ${\beta}$,
we no longer need those $i_k$ that sit adjacent to a vacancy. After dropping
these $i_k$ from  $(i_1,i_2,\cdots,i_L)$, let $(j_1,j_2,\cdots,j_u)$ denote
the remaining sequence of length $u$. This sequence can be coded by means
of an integer $\sigma(j_1,j_2,\cdots,j_u)$ in the range
$1\leq \sigma \leq a_u$. The actual values of $\sigma$ and of $a_u$ are 
given in Ref.~\onlinecite{BN}.
Let \[ A({\beta})=\Sigma_{{\alpha}=1}^{{\beta}-1}
 a_{u({\alpha})}\] be the number of connectivities whose binary
vacancy number is smaller than
${\beta}$, where $u({\alpha})$ is the number of dangling edges
which are not adjacent to a vacant face in the face configuration
${\alpha}$.  Then, the integer that codes the connectivity with
vacancies is
\begin{equation}
\gamma(i_1,i_2,\cdots,i_L)=A({\beta})+\sigma(j_1,j_2,\cdots,j_u)\,.
\end{equation}
A decoding algorithm, that constructs a sequence $(i_1,i_2,\cdots,i_L)$
given the integer $\gamma$, was constructed using similar methods.

\subsection{Numerical calculations}
Several eigenvalues of translationally invariant (zero-momentum)
eigenstates of the transfer matrix were computed for a 
limited range of system sizes $L \leq 16$, as follows:
\begin{enumerate}
\item
The largest eigenvalue $\Lambda^{(0)}_{L}$ in the `even sector', which
means that the transfer matrix operates in the space of connectivities
whose dangling bonds occur only in connected pairs.
\item
The second largest eigenvalue $\Lambda^{(1)}_{L}$ in the same
sector.
\item
The largest eigenvalue $\Lambda^{(2)}_{L}$ in the `odd sector',
which means that the transfer matrix operates in the space of 
connectivities with, apart from dangling pairs of bonds, precisely 
one dangling bond that is single.
\item
The largest eigenvalue $\Lambda^{(3)}_{L}$ in the even sector of
the transfer matrix of a model with a `seam'. The seam modifies one
row of bonds. These bonds are perpendicular to the axis, while the
row itself is parallel to the axis. All edges of this seam contribute
a factor $-1$ to the Boltzmann weight, if covered by a loop segment.
In actual calculations, this is realized by changing the sign of
some of the vertex weights of Fig.~\ref{v2} and Eq.~(\ref{wvwv}),
for those vertices that are immediately to the left of the seam.
\end{enumerate}
The finite-size data for the largest eigenvalue $\Lambda^{(0)}_{L}$
determine the free energy density, from which we estimated the
conformal anomaly $c$ \cite{BCN,Aff}. The ratio
$\Lambda^{(1)}_{L}/\Lambda^{(0)}_{L}$ defines the correlation length
of the energy-energy correlation function. Using Cardy's conformal
mapping \cite{Cardy-xi}
of an infinite cylinder on the infinite plane, one can thus
estimate the temperature dimension $X_t$. Similarly,
$\Lambda^{(2)}_{L}/\Lambda^{(0)}_{L}$
is used to find the magnetic dimension $X_h$. Finally
the ratio $\Lambda^{(3)}_{L}/\Lambda^{(0)}_{L}$ yields the
so called interface exponent $X_{\rm int}$. All of the quantities
$c$, $X_t$, $X_h$, and $X_{\rm int}$ were already described exactly
as a function of $n$, and verified numerically, see Ref.~\onlinecite{GNB}
and references therein. The present numerical analysis is aimed at 
confirming that the present model describes the tricritical
O($n$) model. 
The numerical analysis follows basically the lines of Refs.~\onlinecite{BN}
and \onlinecite{BN82}; see also \onlinecite{FSS}. The final estimates are 
listed in Tabs.~\ref{tab_1} and \ref{tab_2}. They agree convincingly
with the analytic expressions listed in Sec.~\ref{up} whose values
are also included in the tables.
\begin{table}
\caption{Conformal anomaly $c$ and interface critical dimension $X_m$ as
determined from
the transfer-matrix calculations described in the text. Estimated error
margins in the last decimal place are given in parentheses. The numerical
results are indicated by `(num)'. For comparison, we include theoretical
values.
}
\label{tab_1}
\begin{center}
\begin{tabular}{||l|l|l|l|l||}
\hline
 $n$ &$c$ (num)     &$c$ (exact)  &$X_m$ (num)   &$X_m$ (exact) \\
\hline
$-2.0 $&$-0.99155$ (1)&$-0.9915599$&$-0.20179901 $ (1)&$-0.201799000$\\
$-1.75$&$-0.91099$ (1)&$-0.9109986$&$-0.17697229 $ (2)&$-0.176972272$\\
$-1.50$&$-0.81973$ (1)&$-0.8197365$&$-0.15164470 $ (2)&$-0.151644706$\\
$-1.25$&$-0.71646$ (1)&$-0.7164556$&$-0.1259301  $ (1)&$-0.125930086$\\
$-1.00$&$-0.59999$ (1)&$-6/10     $&$-0.100000000$ (1)&$-1/10       $\\
$-0.75$&$-0.46962$ (1)&$-0.4696195$&$-0.07409548 $ (2)&$-0.074095457$\\
$-0.50$&$-0.32528$ (1)&$-0.3252829$&$-0.048531921$ (1)&$-0.048531921$\\
$-0.25$&$-0.16799$ (1)&$-0.1679953$&$-0.023691688$ (1)&$-0.023691689$\\
~~0    &~~0           &~~0         &~~0               &~~0           \\
~~0.25 &~~0.175264 (1)&~~0.1752630 &~~0.02211104  (1) &~~0.0221110351\\
~~0.50 &~~0.353480 (1)&~~0.3534792 &~~0.042235700 (1) &~~0.0422356998\\
~~0.75 &~~0.529949 (1)&~~0.5299489 &~~0.060000362 (1) &~~0.0600003616\\
~~1.00 &~~0.700000 (1)&~~7/10      &~~0.07500000  (1) &~~3/40        \\
~~1.25 &~~0.85897  (1)&~~0.8589769 &~~0.086505216 (2) &~~0.0865052157\\
~~1.50 &~~1.00000  (1)&~~1         &~~0.08801923  (5) &~~0.0880192310\\
\hline
\end{tabular}
\end{center}
\end{table}
\begin{table}
\caption{Temperature critical dimension $X_t$ and magnetic dimension
$X_h$ as determined from
the transfer-matrix calculations described in the text. Estimated error
margins in the last decimal place are given in parentheses. The numerical
results are indicated by `(num)'. For comparison, we include theoretical
values.
}
\label{tab_2}
\begin{center}
\begin{tabular}{||l|l|l|l|l||}
\hline
   $n$ &$X_t$ (num)   &$X_t$ (exact)&$X_h$ (num)   &$X_h$ (exact)\\
\hline
$-2.0 $& ~~~~------     & ~~~~------  &$-0.0951628  $ (1)& $-0.0951627339$\\
$-1.50$& ~~~~------     & 0.709784688 &$-0.0876432  $ (1)& $-0.0876431495$\\
$-1.25$& 0.4814737  (2) & 0.481473928 &$-0.0790909  $ (1)& $-0.0790908776$\\
$-1.00$& 0.39999999 (1) & 2/5         &$-0.05833333 $ (1)& $-7/120       $\\
$-0.75$& 0.3446680  (2) & 0.344668096 &$-0.045889544$ (1)& $-0.0458895426$\\
$-0.50$& 0.3039307  (2) & 0.303930873 &$-0.031982842$ (2)& $-0.0319828413$\\
~~0.00 & 0.2500000  (1) & 1/4         &~~0               &~~0             \\
~~0.25 & 0.2324956  (1) & 0.232495729 &~~0.017729518 (1) &~~0.0177295181  \\
~~0.50 & 0.2192386  (1) & 0.219238626 &~~0.03627658  (1) &~~0.0362765827  \\
~~0.75 & 0.2088742  (1) & 0.208874121 &~~0.05539746  (1) &~~0.0553974632  \\
~~1.00 & 0.20000001 (1) & 1/5         &~~0.07500000  (1) &~~3/40          \\
~~1.25 & 0.19068002 (1) & 0.190680043 &~~0.09549715  (1) &~~0.0954971419  \\
~~1.50 & 0.16844985 (5) & 0.168449854 &~~0.125000000 (1) &~~1/8           \\
\hline
\end{tabular}
\end{center}
\end{table}

\section{Discussion}
\label{dis}
The present tricritical O($n$) model appears to belong to the same
universality class as a loop model defined in Ref.~\onlinecite{NWB}.
The latter model was defined as the surrounding loop model of
the critical $q$-state random cluster model on the square lattice.
It is possible to apply the same method as used above, namely to use
the ADE interpretation and to restore the loops except those surrounding
the vacancies of type $a$ and $d$, to the latter loop model. We have
chosen the present formulation, based on the low-temperature O($n$)
model of branch 2 defined in Refs.~\onlinecite{BNW} and \onlinecite{BN}.
This is more natural
in the sense that it allows for sites that are neither visited by a loop,
nor adjacent to a vacancy. The relations between the 
various models, as constructed and listed in Secs.~\ref{smap} and
\ref{exs}, are summarized by
\begin{equation}
Z_{\rm spin} \leftrightarrow Z_{\rm dense~loop} \leftrightarrow Z_{\rm ADE}
\leftrightarrow Z_{\rm loop+vac.} \leftrightarrow Z_{\rm dilute ~ spin}\;.
\label{map1}
\end{equation}
The last step follows from the general equivalence formulated in
Sec.~\ref{smap}.

As found in Ref.~\onlinecite{GNB}, the introduction of vacant faces in
the honeycomb O($n$) model leads to tricriticality when the fugacity
of the vacancies is sufficiently large. No vacancy-vacancy couplings
were introduced. Numerical work on the square lattice O($n$) model 
for $n=1$ revealed a peculiar difference with the honeycomb O($n$) 
model. No tricritical point was found when vacancy-vacancy couplings
are absent. Instead, a multicritical point resembling that of branch
3 of Ref.~\onlinecite{BN} was found. The physical interpretation of this
multicritical point is that the O($n$) critical line merges with an
Ising critical line, where auxiliary variables in the form of dual
Ising spins undergo a phase transition. A qualitative difference
with the model described by the vertex weights of Eq.~(\ref{wvwv})
is that the vacancies attract each other in the latter model. 

A comparison of the numerical results for the present model with those
for the tricritical honeycomb model O($n$) studied in Ref.~\onlinecite{GNB}
shows a conspicuous difference in the estimated accuracies. This
difference can be explained from the way in which the two different
sets of tricritical points were found. For the honeycomb lattice O($n$)
model of Ref.~\onlinecite{GNB}, the tricritical points were determined
numerically in a small parameter space. From the perspective of the
renormalization theory, this procedure yields a rather arbitrary
tricritical point in the sense that the irrelevant fields are non-zero
in general, and thus introduce corrections to scaling.  In contrast,
the exact equivalence of the present tricritical square-lattice O($n$)
model with the O($n$) low-temperature branch indicates that the leading
irrelevant field vanishes, since the equivalent O($n$) low-temperature 
branch is characterized by the vanishing of its irrelevant temperature
field. As a result the corrections to scaling are suppressed, and the
apparent finite-size convergence improves drastically.

Since it is widely believed that the universal parameters describing
the critical state are determined by the symmetry of the model,
the dimensionality, and the range of interaction, it seems plausible
that the tricritical model presented above serves as a representative
of the generic O($n$) universality in two dimensions. Indeed the 
spin-spin interactions defined in Sec.~\ref{smap} contain only
scalar products, which satisfy the O($n$) symmetry.

However, in this case the O($n$) symmetry of the spin model, is not a
secure guide for the universality class.  This is because, like in the
dense loop phase of the pure O($n$) model, intersections are relevant
\cite{N,BN2}. The same applies to the tricritical point reported here:
the exponent associated with crossing loops is the same as that of
cubic symmetry breaking.  Recently, Jacobsen et al.~\cite{jrs} proposed
that the low-temperature phase of the generic O($n$) model is described
by the intersecting loop model proposed in \cite{MNR} and since called
Brauer model~\cite{dGN}.

It is interesting to note that the mappings described in Sec.~\ref{exs}
can also be applied to the critical `branch 1' \cite{BN,BNW,WPSN} of
the square-lattice O($n$) model. Just as branch 1 is the analytic 
continuation of branch 2, we can continue the tricritical branch
through the `end point' $n=3/2$, $g=1$ to $g>1$. The weights for
branch 1 are also given by Eq.~(\ref{clt}), but instead with 
$\pi/2<\theta<\pi$. The relation
between $n$ and $n'$ remains the same, but the vertex weights as
specified by Eq.~(\ref{clt}) change, and the relation between the
Coulomb gas coupling and the conformal classification parameter $m$
is no longer $g=m/(m+1)$ but becomes $g=(m+1)/m \geq 1$, while it
relates to $n'$ as $n'=2\cos(\pi/m)$ (see e.g. Ref.~\onlinecite{BN}). 
For $n=1$ or $n'=(1+\sqrt{5})/2$ one thus
finds a higher critical Ising model that is to be compared with the
$m=5$ model in the series of Andrews et al.~\cite{ABF}. For $n=0$ or
$n'=1$ the model displays an Ising-like critical point. The resulting
branch of multicritical points can thus be seen as a generalization
of the $m=5$ Ising-like model for $n=1$ to continuous values of $n$,
i.e. the point where the tricritical point itself turns first-order.
\vspace{5mm}
\acknowledgments
This research is supported by the NSFC under Grant \#10675021, by
the Beijing Normal University through a grant as well as support from
its HSCC (High Performance Scientific Computing Center), and, in part,
by the Lorentz Fund.
We thank Youjin Deng for some valuable discussions.

\end{document}